# One-dimensional optomagnonic microcavities for selective excitation of perpendicular standing spin waves

V.A. Ozerov[1,2,3,a], D.A. Sylgacheva[2,4], M.A. Kozhaev[1,2], T. Mikhailova[1], V.N. Berzhansky[1], Mehri Hamidi[5], A.K. Zvezdin[1,2], V.I. Belotelov[1,2,4]

[1]*Vernadsky Crimean Federal University, Simferopol, Russia*
[2]*Russian Quantum Center, Skolkovo, Moscow, Russia*
[3]*Moscow Institute of Physics and Technology (National Research University), Dolgoprudny, Russia*
[4]*Faculty of Physics, Lomonosov Moscow State University, Moscow, Russia*
[5]*Laser and Plasma Research Institute, Shahid Beheshti University, Evin Tehran, Iran*

[a]Corresponding author: vladislav.ozerov@skoltech.ru

**Abstract.** Here we propose a method of the excitation of perpendicular standing spin waves (PSSWs) of different orders in an optomagnonic microcavity by ultrashort laser pulses. The microcavity is formed by a magnetic dielectric film surrounded by dielectric non-magnetic Bragg mirrors. Optical cavity modes in the magnetic layer provide concentration and strongly non-uniform distribution of the optical power over the layer thickness and therefore induce the effective field of the inverse Faraday effect also spatially non-uniform. It results in excitation of PSSWs. PSSWs whose wavevector is closest to the wavevector characterizing distribution of the inverse Faraday effect field are excited most efficiently. Consequently, a key advantage of this approach is a selectivity of the PSSW excitation which allows to launch PSSWs of required orders only. All-optical operation of the optomagnonic cavities opens new possibilities for their applications for quantum technologies.

## I. INTRODUCTION

The demand for energy efficient data processing has put magnonics among actively studied areas of modern physics [1–3]. Propagation of magnons or spin waves in magnetically ordered media doesn't involve charge transport and therefore heating losses are significantly reduced. Moreover, spin excitations are very promising for quantum technologies [3–6]. In particular, quantum transitions between magnonic Bose-Einstein condensate states might be utilized for the single-qubit gates [6]. Apart from that, magnons can mediate the microwave-to-optical conversion necessary for quantum memories [7,8] and optical interfacing of the superconducting qubits [4,5].

In most of these applications optomagnonic cavities play a crucial role since they provide spatial localization of both spin and electromagnetic waves at the same region and consequently enhance the coupling between them which results in a higher conversion efficiency between microwave and optical photons. Moreover, in the optomagnonic cavities the strong-coupling regime might be achieved when photons and magnons hybridize, forming a quasiparticle called magnon-polariton [9,10].

Optomagnonic cavities of different architecture were considered, such as photomagnonic crystals [11,12], planar waveguides [8], yttrium iron-garnet (YIG) spheres [13–17], and microcavities [18–24]. As for the submillimeter YIG spheres, they support optical whispering gallery modes [13–15] and Mie resonances [25]. Consequently, the YIG spheres may be used, for example, for realizing magnon-assisted photon transitions between the Mie modes, when the triple resonance condition is met.

On the other hand, the optomagnonic microcavities, consisting of ultrathin magnetic dielectric films sandwiched in between the dielectric multilayer Bragg mirrors, are advantageous for light localization in the magnetic films due to the Fabry-Perot resonances (optical cavity modes) spectrally lying within the photonic band gap of the Bragg mirrors [18]. It was proposed theoretically, that the perpendicular standing spin waves (PSSW), also resonating in the microcavity, interact efficiently with light leading to the strong magneto-optical interaction beyond the linear regime and providing enhanced modulation of light through multimagnon absorption and emission processes [18–20].

However, the optomagnonic microcavities might be considered the opposite way around. i.e. to influence spins by light. Actually, launch of spin precession and spin waves as well as magnetization switching have been successfully accomplished in different magnetic crystals and films by femtosecond laser pulses [26–35]. Recently, advantages of optical resonances in nanophotonic structures were taken providing localized and enhanced excitation of spin waves [36–43]. In this respect photonic crystals with magnetic layers are very promising since they provide a significant localization of optical power within the magnetic layer which increases the magneto-optical interaction [44–46]. Experiments with one dimensional photonic crystal cavity demonstrated that the optical cavity modes enhance the inverse Faraday effect (IFE) which was found by a pronounced increase of the spin wave amplitude excited by ultrashort laser pulses [24]. In Ref.[24] the backward volume magnetostatic spin waves were observed only.

It would be advantageous to excite PSSWs in similar manner. Indeed, PSSW are on prime demand for the aforementioned quantum applications and the problem of PSSW efficient excitation still seeks solution. The previous approaches for PSSW generation were based on application of microwave field using special antennas [47], planar microwave waveguides [48], exchange torques [49] etc. Depending on the way of excitation, several magnonic modes can be triggered, however, the most efficiently excited one being the fundamental uniform precession zero mode ($n = 0$) since generation of the higher-order modes ($n > 0$) requires a non-uniform dynamical field across the thickness of a magnetic film, which is very challenging to obtain using microwave antennas. At the same time, due to the restrictions imposed by the size of the microwave antennas on the excited spin wave wavelength, magnetostatic spin wave modes ($n = 0$) in such devices did not exceed the frequencies of few GHz at moderate magnetic fields. The other drawback of the microwave approach is its very limited tunability. Spectrum of PSSW is fully determined by the microwave antenna shape and position which are fixed. Next, microwave stimulus is difficult to be localized and generally influences on rather large area of a magnet limiting miniaturization level of the magnonic devices. Finally, it is very difficult to excite only a single PSSW harmonics by microwaves.

In this respect optical means holds a big promise since optical field distribution inside a magnetic film might be significantly localized up to Rayleigh limit and even beyond. Moreover, light impact can be tuned by changing wavelength or incidence angle.

Recently, M. Deb et al. [50] have shown that ultrafast laser excitation leads to the non-uniform modification of magnetic anisotropy through the film thickness, that, in turn, results in excitation of 15 GHz PSSWs at low external fields. However, the origin of the demonstrated high frequency oscillations is of the thermal character, thus, it would lead to extra heat losses. Apart from that, optical field distribution inside a single magnetic film can't be notably altered by varying incident light parameters.

In this work we propose and demonstrate a novel method for excitation of PSSWs of different orders in the optomagnonic microcavities by ultrashort laser pulses. The principle is the following: a circularly polarized laser pulse passing through the ferromagnetic film inside the microcavity effectively generates a magnetic field directed along light wavevector in accordance to the inverse Faraday effect (IFE). We show, that light pulses allow to excite different orders of the PSSWs modes selectively, i.e. to excite only required spin waves harmonics. This is a unique feature of the optical approach.

The paper is organized as follows. Sec.I gives the introduction to the problem. Sec.II states the problem, defines the configuration of the optomagnonic structure and the orientation of all the vectors. In Sec.III all the necessary equations of spin dynamics are given, the formulas for the PSSW eigenmodes, their wavevectors $k_n$ and the excitation amplitudes of modes $A_n$ are obtained. In Sec.IV several configurations of optomagnonic structures are considered, which can be used in a real experiment to excite PSSWs. The corresponding IFE-field distributions for the given structures are modelled. In Sec.V the excitation amplitudes $A_n$ are calculated for the different IFE-field distributions, presented in Sec.IV. It is shown, that only single modes are excited in a ferromagnetic film, but not the whole spectrum of modes. Sec.VI gives a short summary of the obtained results and outlines the prospects of the future possible implementations of the suggested method of optical excitation of PSSWs.

## II. PROBLEM STATEMENT

When a magnetic medium is illuminated by a circularly polarized light its spin angular momentum is transferred to the spin system of the medium through the stimulated Raman scattering, the phenomenon usually referred to as the inverse Faraday effect. This process can be described in terms of the effective magnetic field $\mathbf{H}_{\text{IFE}}$ induced by laser pulses: $\mathbf{H}_{\text{IFE}} = -\frac{g}{16\pi}\text{Im}\{[\mathbf{E} \times \mathbf{E}^*]\}$ [51,52], where $\mathbf{E}$ is electric field of light inside the magnetic film and $g$ is the magnetooptical gyration constant. For the normal incidence of circularly polarized light $\mathbf{H}_{\text{IFE}}$ is directed parallel to the light wavevector, i.e. along the normal (z-axis in Fig. 1). The induced magnetic field exists in the medium only during pulse propagation and deflects the magnetization $\mathbf{M}$ from its equilibrium orientation. If the external magnetic field $\mathbf{H}$ lies in the sample plane and crystal anisotropy can be neglected then the equilibrium position of $\mathbf{M}$ coincides with $\mathbf{H}$ (x-axis in Fig. 1). After its deflection the magnetization starts to precess around $\mathbf{H}$. Due to the exchange and magnetic dipole-dipole interaction different kinds of spin waves can be excited in the magnetic medium. Generally, magnetostatic spin waves propagating away from the illuminated area and exchange perpendicular standing spin waves should appear. However, if the laser beam diameter is large enough the magnetostatic spin waves are excited with wavevector close to zero and can be considered as a uniform precession in lateral plane (x-y plane in Fig. 1). On the other hand, the standing spin waves have non uniform oscillation distribution along the film thickness and thus require non-uniform excitation of the external stimulus, i.e. $\mathbf{H}_{\text{IFE}}(z)$, which can be induced in the magnetic film placed in a magnetic microcavity.

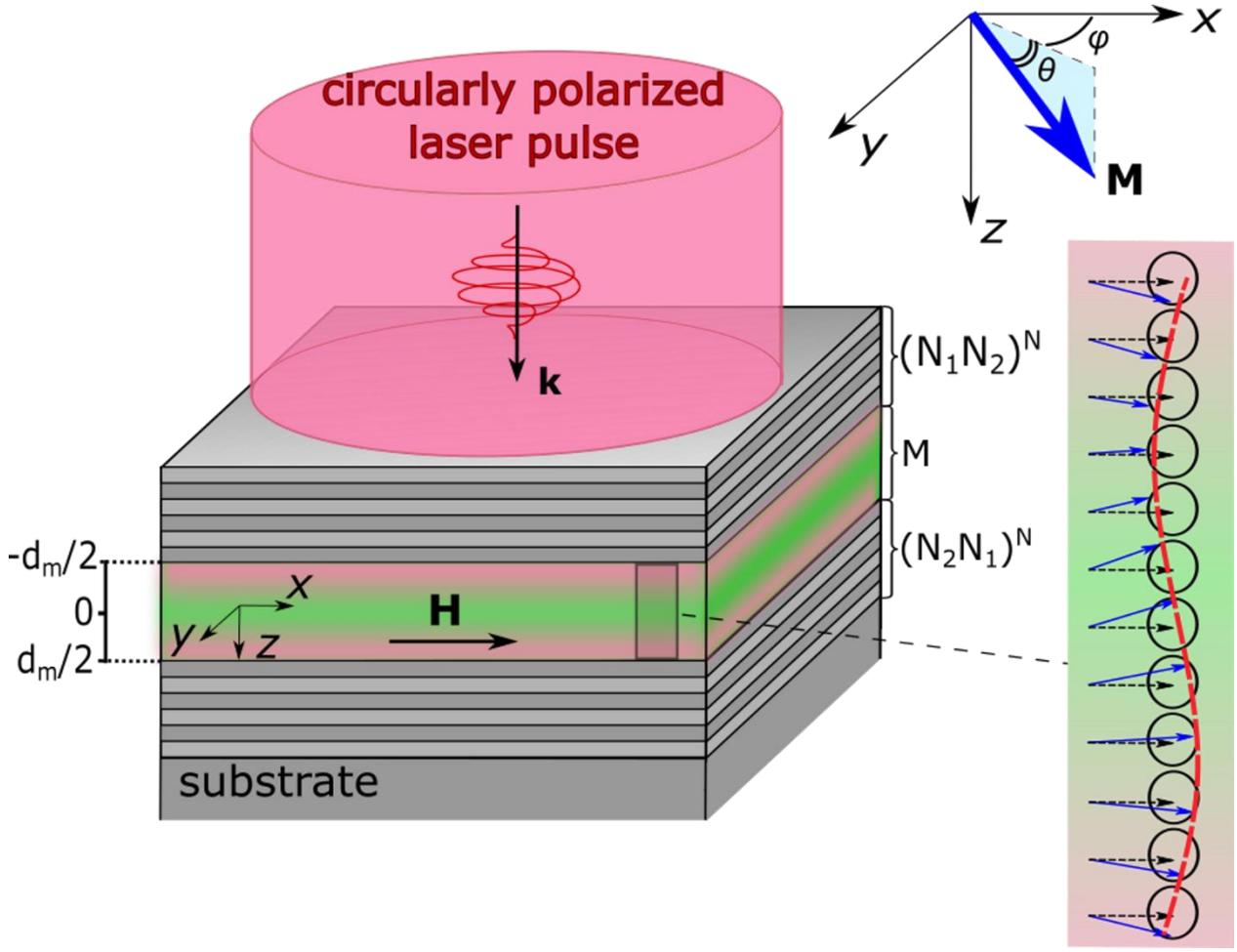

FIG. 1. The sample is an optomagnonic structure formed by the magnetic film M sandwiched in between two nonmagnetic Bragg mirrors formed by several pairs of the dielectric layers $N_1$ and $N_2$. The circularly polarized pump (pink cylinder) induces some distribution of the IFE-field which, in its turn, excites spin dynamics in the magnetic film. One of the possible distributions of the effective magnetic field is shown on the front side of the sample by pink and green color (pink corresponds to the maximal field). The right inset shows schematically excited perpendicular standing spin waves. The origin of the coordinate system (shown on right up) is in the middle of the magnetic film.

For the circularly polarized laser pulse the induced electric field has a form: $\mathbf{E}_\pm(t,z) = \tilde{f}(t)E_0(z)\boldsymbol{\sigma}_\pm e^{-i\omega t}$, where $\tilde{f}(t)$ is envelope function of the laser pulse in the microcavity, $E_0(z)$ is the normalized absolute value of the electric field of light responsible for its spatial distribution, $\boldsymbol{\sigma}_\pm$ is the unit vector of circular polarization in x-y plane, $\omega$ is frequency of light. In this case $\mathbf{H}_{\text{IFE}}$ is given by:

$$\mathbf{H}_{\text{IFE}}(t,z) = \pm \mathbf{e}_z f(t) h(z), \quad (1)$$

where $h(z) = \frac{g}{16\pi}|E_0(z)|^2$, $f(t) = \tilde{f}^2(t)$, $\mathbf{e}_z$ is unit vector along z-axis. The envelope function $f(t)$ is non-zero only during the time $\Delta t$ of laser pulse propagation ($f(t) \neq 0$ for $0 \leq t \leq \Delta t$). The functions $f(t)$, $h(z)$ (and therefore $\tilde{f}(t), E_0(z)$ ) will be normalized to satisfy the condition $\overline{|\mathbf{H}_{\text{IFE}}|}(z) = \frac{1}{\Delta t}\int_0^{\Delta t}|\mathbf{H}_{\text{IFE}}(t,z)|\,dt = h(z)$, meaning that $h(z)$ is the average value of the IFE-field during the pulse propagation. In this case, $\int_0^{\Delta t} f(t)\,dt = \Delta t$. As for $E_0(z)$, it plays the role of the root mean square of the induced electric field.

As the duration of femtosecond laser pulses is much smaller than the precession period of spins in the system ($\Delta t \ll T$), then the influence of $\mathbf{H}_{\text{IFE}}$ can be reduced to the determination of initial conditions of magnetization $\mathbf{M}$ (as will be shown in Eqs. (3), (4) ). Further dynamics (for $t > \Delta t$) is determined by properties of the system itself, without taking into account IFE-field $\mathbf{H}_{\text{IFE}}$.

The optomagnonic microcavity considered here consists of a subwavelength-thick magnetic layer surrounded by non-magnetic Bragg mirrors (Fig. 1). On the one side, such a thin magnetic layer sustains a comb of PSSWs of different orders at GHz frequencies. On the other hand, if light passes through this structure, a distribution of its optical power is strongly non-uniform in space. In particular, the optomagnonic structure is a kind of 1D photonic crystal having a specific transmission spectrum with photonic band gap and a microresonator peak in the band gap. If light wavelength is tuned, for example, at the microresonator peak or at the edge of the photonic band gap, then distribution of the optical intensity inside the magnetic layer has pronounced maxima and minima similar to the nodes and anti-nodes of a standing wave but their locations are different. It might be used for optical excitation of different PSSW modes.

To solve the problem of the PSSW excitation we, firstly, derive expressions describing the magnetization dynamics excited at an arbitrary function of $h(z)$ and then investigate what kind of different distributions of $h(z)$ can be implemented in the magnetophotonic microcavity under illumination by light. Finally, we will use these distributions to calculate spectrum of PSSWs which can be excited in a real microcavity sample.

### III. PSSWS EXCITED BY A SPATIALLY NON-UNIFORM INSTANT STIMULUS

Magnetization dynamics launched by ultrashort laser pulses can be investigated on the basis of the Landau-Lifshitz-Gilbert equation:

$$\frac{d\mathbf{M}}{dt} = -\gamma[\mathbf{M} \times \mathbf{H}_{eff}] + \frac{\alpha}{|\mathbf{M}|}\left[\mathbf{M} \times \frac{d\mathbf{M}}{dt}\right]. \quad (2)$$

Here the effective magnetic field acting on magnetization is $\mathbf{H}_{eff} = \mathbf{H} + 4\pi\mathbf{M} - \frac{2K_U}{M}\mathbf{e}_z + A\Delta\mathbf{M} + \mathbf{H}_{IFE}$, where $K_U$ is uniaxial anisotropy constant, $A$ is exchange constant, $\gamma$ is gyromagnetic ratio and $\alpha$ is Gilbert damping constant considered to be small ($\alpha \sim 10^{-3}$). In the spherical coordinate system (Fig. 1) we have $M_y = |\mathbf{M}|\cos\theta\sin\varphi$; $M_z = |\mathbf{M}|\sin\theta$ and Eq. (2) is rewritten as:

$$\begin{cases} \dot{\theta} = \alpha\dot{\varphi} + \gamma H\varphi - \gamma AM\varphi'', \\ \dot{\varphi} = -\alpha\dot{\theta} - \gamma\widetilde{H}\theta + \gamma AM\theta'' + \gamma f(t)h(z). \end{cases} \quad (3)$$

Here we assumed that the precession angle is small ($\theta \ll 1$, $\varphi \ll 1$), and the derivatives are denoted by $\theta'' = \frac{\partial^2\theta}{\partial z^2}$, $\varphi'' = \frac{\partial^2\varphi}{\partial z^2}$, $\dot{\theta} = \frac{\partial\theta}{\partial t}$, $\dot{\varphi} = \frac{\partial\varphi}{\partial t}$. For convenience, we introduced the notation $\widetilde{H} = H + 4\pi M - \frac{2K_U}{M}$, where $H = |\mathbf{H}|$, $M = |\mathbf{M}|$.

The IFE-term $\gamma f(t)h(z)$ in Eq. (3) is non-zero only during the small time $\Delta t$, meaning that it is responsible only for the establishing of the initial conditions for $\theta, \varphi$. If we integrate Eq. (3) by $t$ from 0 to $\Delta t$ using $\int_0^{\Delta t} f(t)\,dt = \Delta t$, we will find the initial conditions for $\theta, \varphi$ after the instant stimulus of a laser pump:

$$\begin{cases} \varphi(z, t = \Delta t) = \gamma\Delta t \cdot h(z), \\ \theta(z, t = \Delta t) = \alpha\gamma\Delta t \cdot h(z) \cong 0. \end{cases} \quad (4)$$

Besides, boundary conditions should also be taken into consideration. As shown in [53], for the given configuration the boundary conditions take the form:

$$\begin{cases} \theta' \mp \xi\theta = 0; & z = \pm\frac{d}{2}, \\ \varphi' = 0; & z = \pm\frac{d}{2}. \end{cases} \quad (5)$$

Here $\xi$ is a pinning parameter, originating from the surface anisotropy, $d$ is the thickness of the magnetic film. Parameter $\xi$ may be expressed in terms of the surface anisotropy parameter $K_s$ as follows: $\xi = \frac{2K_s}{AM^2}$. We are interested in the value of the dimensionless product $\xi d$, as the eigenmodes of the system strongly depend on it. For the further calculations we take $K_s = 2.9 \cdot 10^{-2}\frac{erg}{cm^2}$ leading to $\xi d = 0.5, 1.0, 1.5$ (for $d = 69, 138, 207$ nm, respectively), which is a realistic estimation [54].

Eq. (3) together with Eqs. (4) and (5) fully formulates the Cauchy differential equation problem. The solution of Eq. (3) has the form of decaying harmonic oscillations $\theta(z,t), \varphi(z,t) \sim e^{i(kz-\omega t)-\lambda t}$. After substitution, we can determine the expressions for the frequency $\omega$ and for the damping parameter $\lambda$:

$$\begin{cases} \lambda = \alpha\gamma\left[\dfrac{H+\tilde{H}}{2}+AMk^2\right], \\ \omega^2 = \gamma^2(H+AMk^2)(\tilde{H}+AMk^2)-\lambda^2 = \omega_0^2-\lambda^2 \approx \omega_0^2. \end{cases} \qquad (6)$$

Two values of wavevector $k$ ($k_+$ and $k_-$) correspond to the given value of the frequency $\omega$ [Appendix (A1), (A2)]. They are linked together by the ratios [Appendix (A3.a)–(A3.b)]. The first one ($k_+$) can be either real ($k_+ = k$) or imaginary ($k_+ = i\chi_+$), depending on $\omega$ [see App.], and the second one ($k_-$) is always imaginary ($k_- = i\chi$). As it will be shown later, modes with imaginary $k_\pm$ are responsible for hyperbolic (surface) terms of the modes, and can't be neglected.

Since the excited PSSW oscillations can be detected through the Faraday effect, which is sensitive to the normal component of the magnetization, we will describe the PSSWs by the $\theta$ angle. There are two types of the PSSW modes $\theta_n(z,t) = \theta_n(t) \cdot \theta_n(z)$:

$$\theta_n(z,t) = e^{-\lambda_n t}\sin\omega_n t \cdot \begin{cases} \cos k_n z + b_{\theta,n} B_{\chi,n}\cosh\chi_n z, & n = 2,4,6,\ldots \\ \sin k_n z + b_{\theta,n} B_{\chi,n}\sinh\chi_n z, & n = (1),3,5,\ldots \end{cases}, \qquad (7a)$$

$$\theta_n(z,t) = e^{-\lambda_n t}\sin\omega_n t \cdot \begin{cases} \cosh\chi_{+,n} z + \tilde{b}_{\theta,n}\tilde{B}_{\chi,n}\cosh\chi_n z, & n = 0 \\ \sinh\chi_{+,n} z + \tilde{b}_{\theta,n}\tilde{B}_{\chi,n}\sinh\chi_n z, & n = (1) \end{cases}. \qquad (7b)$$

Here, the modes symmetric with respect to the film center correspond to even $n$, the antisymmetric modes – to the odd $n$. Note, that the mode with $n = 1$ may have a form of Eq. (7a) or Eq. (7b) depending on the product $\xi d$: if $\xi d < 6 * \left(1 + \dfrac{4\tanh\left[\sqrt{\tfrac{\tilde{H}+H}{AM}}\tfrac{d}{2}\right]}{\sqrt{\tfrac{\tilde{H}+H}{AM}}\,d}\right)^{-1}$, then Eq. (7a) is valid. For the further analysis we consider $\xi d = 0.5 - 1.5$, thus, $n = 1$ mode will take a form of Eq. (7a) [see Appendix (A17)–(A19)].

It should be noted that for small values of $\xi d$ hyperbolic terms in Eq. (7a) are relatively small and these modes can be considered as quasi-harmonic. The expressions for $b_{\theta,n}, \tilde{b}_{\theta,n}, B_{\chi,n}, \tilde{B}_{\chi,n}$ coefficients, which are responsible for the hyperbolic terms, are given in Appendix [see Eqs. (A7), (A8), (A14), (A15)].

Wavevector of the «quasi-harmonic» modes by Eq. (7a) ($k_+^2 > 0$) is described as [Appendix (A9)–(A11)]:

$$k_n = \frac{\pi n}{d} - \frac{2\xi^*}{\pi n}, \qquad n = (1),2,3,4,\ldots. \qquad (8)$$

Here, odd numbers of modes ($n = 1,3,5,\ldots$) correspond to the antisymmetric solutions, and the even numbers ($n = 2,4,6,\ldots$) correspond to the symmetric solutions. The mode $n = 1$ is taken in parentheses, because it may not satisfy the condition $k_+^2 > 0$, depending on product $\xi d$. However, for $\xi d = 0.5 - 1.5$ the Eq. (8) is applicable to it. As for the wavevector $\chi_n$ in Eq. (7.a), it is linked together with $k_n$ by the ratio [Appendix (A3.a)] and, thus, can be easily found.

We should also analyze the case, when $k_+^2 < 0$ which might take place [see App.] for the first two modes ($n = 0,1$). In this case the modes are described by Eq. (7b). We will call them «hyperbolic» modes, as they are expressed through the sum of hyperbolic functions and have imaginary wavevectors $k_{+,n} = i\chi_{+,n}$, $k_{-,n} = i\chi_n$. These wavevectors, which are linked together by the ratio [Appendix (A3.b)], can be found numerically from [Appendix (A16)].

A spatially non-uniform instant stimulus will excite a set of eigenmodes with different amplitudes $A_n$:

$$\theta(z,t) = \sum_{n=0}^{\infty} A_n \cdot \theta_n(z,t). \qquad (9)$$

In a case, when the IFE-field is proportional to the superposition of the harmonic functions $h_s(z) \sim \sin k_s z, \cos k_s z$ (including $k_s = 0$): $h(z) = \sum_s h_s(z)$, the expression for the amplitudes of modes $A_n$ is given by [Appendix (A23)]. Depending on the character of mode $\theta_n(z)$ («quasi-harmonic» or «hyperbolic») we finally have:

$$A_n = \frac{\beta_n \gamma \Delta t}{d}\sum_s \frac{H + AMk_s^2}{\sqrt{(H+AMk_n^2)(\tilde{H}+AMk_n^2)}} \int_{-d/2}^{d/2} h_s(z)\cdot\theta_n(z)\,dz, \qquad n = (1),2,3,\ldots \qquad (10a)$$

$$A_n = \frac{\beta_n \gamma \Delta t}{d} \sum_s \frac{H + AMk_s^2}{\sqrt{(H - AM\chi_{+,n}^2)(\widetilde{H} - AM\chi_{+,n}^2)}} \int_{-d/2}^{d/2} h_s(z) \cdot \theta_n(z) \, dz, \qquad n = 0, (1) \ , \quad (10b)$$

where $\beta_{n=0} = 1$, $\beta_{n \neq 0} = 2$. It should be noted that the amplitudes of different PSSW eigenmodes are mostly defined by the overlapping integral in Eqs. (10a)–(10b). At this, the maximum values of $A_n$ should correspond to the PSSW modes, that have the wavevector $k_n$ close to $k_s$ of the IFE-field.

## IV. THE IFE-FIELD DISTRIBUTION IN THE MICRORESONATOR MAGNETIC LAYER

As we could see from Eqs. (10a)–(10b) the PSSWs spectrum depends on the IFE-field spatial distribution in the magnetic layer. Here we will consider several examples of the IFE-field distribution which can be set in an optomagnonic microcavity. The cavity is originated by a magnetic dielectric layer M of thickness $d_m$ surrounded by the Bragg mirrors (Fig. 1). The Bragg mirrors are composed of $N$ pairs of two nonmagnetic dielectric layers $N_1$ and $N_2$ each of the quarter-wavelength thickness, $\lambda_0/4\sqrt{\varepsilon_i}$, with respect to the central wavelength of the photonic band-gap, $\lambda_0$, of the Bragg mirrors. Here $\varepsilon_i$ is permittivity of the $N_i$-th layer. Therefore, the whole structure is as follows: [substrate / $(N_1/N_2)^N$ / M / $(N_2/N_1)^N$]. For the exemplary calculations we assume the three-pair Bragg mirrors ($N = 3$), and $N_1$ represented by $Ta_2O_5$ and $N_2$ – by $SiO_2$, M is bismuth substituted iron-garnet [55,56] and the substrate is gadolinium gallium garnet. Calculations of the light propagation in the optomagnonic structure were performed by the transfer matrix method [57].

We chose $\lambda_0 = 680$ nm where the magneto-optical figure of merit of iron-garnet is relatively large and start from the magnetic layer of half-wavelength thickness, i.e. $d_m = \lambda_0/2\sqrt{\varepsilon_m}$, where $\varepsilon_m$ is permittivity of the magnetic layer. The transmission spectrum of such structure has a pronounced photonic band-gap in the spectral range from 600 nm to 800 nm with minimum transmission of a few percent (Fig. 2(a)). It becomes possible even for $N = 3$ pairs since optical contrast between $Ta_2O_5$ and $SiO_2$ is quite large. In the center of the photonic band gap there is a transmission peak at $\lambda_0 = 680$ nm corresponding to excitation of the optical cavity mode in the magnetic layer (Fig. 2(b), red curve).

If the structure is illuminated by a circularly polarized laser pulse at $\lambda_0$ to excite the cavity mode, then light intensity oscillates inside the Bragg mirrors and acquires maxima at the both faces of the magnetic layer ($z = \pm d_m/2$). Though zero intensity appears at the magnetic layer center ($z = 0$), most of the pulse energy is still concentrated within the magnetic layer. Optical field induces the IFE-field $\mathbf{H}_{\text{IFE}}$ directed along the pulse wavevector and existing during the pulse propagation. Its distribution $h(z)$ is shown by blue curve in Fig. 2(b). Therefore, the IFE-field spatial distribution is described by cosine function shifted by a constant: $h(z) = \frac{h_0}{2}(1 - \cos k_s z)$ with $k_s = s\pi/d_m$ and $s = 2$.

Passing to a thinner magnetic layer of $d_m = \lambda_0/4\sqrt{\varepsilon_m}$ drastically changes transmittance (Fig. 2(c)) and optical field distribution (Fig. 2(d)) in the photonic structure. Namely, the transmission peak moves away from the photonic band gap and light intensity oscillates and exponentially decays inside the structure (red curve in Fig. 2(d)). Nevertheless, some part of the incident light penetrates through the front Bragg mirror to the magnetic layer and induces the IFE field (blue curve in Fig. 2(d)). On the contrary to the previous case, it is asymmetric having zero at the front side of the magnetic layer ($z = -d_m/2$) and maximum $h_0$ at the back side ($z = d_m/2$). Consequently, $h(z)$ is well described by sine function with added constant: $h(z) = \frac{h_0}{2}(1 + \sin k_s z)$, where $s = 1$.

An increase of the magnetic layer thickness leads to additional maxima and zeros of $h(z)$. Thus, for $d_m = 3\lambda_0/4\sqrt{\varepsilon_m}$: $h(z) = \frac{h_0}{2}(1 - \sin k_s z)$ with $s = 3$ (Fig. 2(f)). As for the quarter-wavelength magnetic layer the transmission peak is away from the photonic band gap (Fig. 2(e)) and light intensity decays inside the photonic stack. In fact, optical resonance in the magnetic layer appears for the thicknesses around $d_m = s\lambda_0/4\sqrt{\varepsilon_m}$ with even $s$ and is absent for odd $s$.

Consequently, change of the magnetic film thickness provides different spatial distribution of the IFE field and should excite various spin wave dynamics. Furthermore, different IFE-field distributions are attainable for the same optomagnonic structure if incident light wavelength is altered. For example, decrease of the light wavelength to $\lambda_0 = 580$ nm for the structure with the quarter-wavelength magnetic layer corresponds to the short-wavelength edge of the photonic band gap (Fig. 2(a)) and maximum and zero of $h(z)$ become located asymmetrically with respect to the magnetic layer center (Fig. 2(g)). Similar situation appears for the illumination at the transmission band, i.e. at $\lambda_0 = 840$ nm (Fig. 2(h)). It broadens capabilities of the optical approach for the excitation of the PSSWs. In the following section we will analyze PSSW spectra excited in the different scenarios.

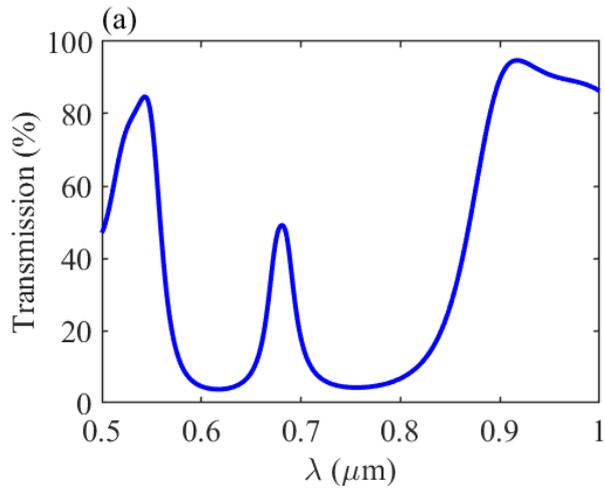
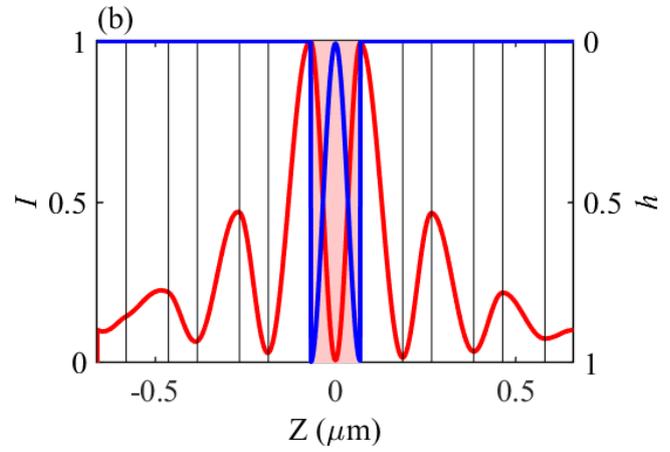
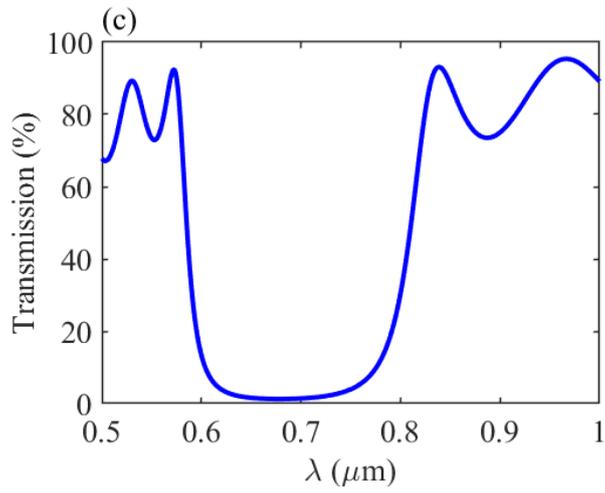
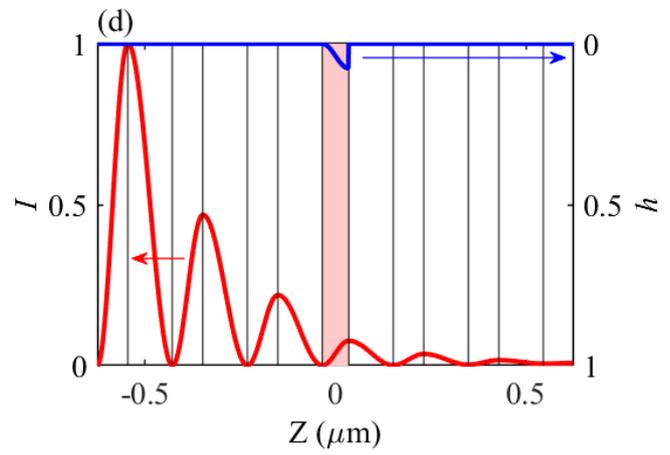
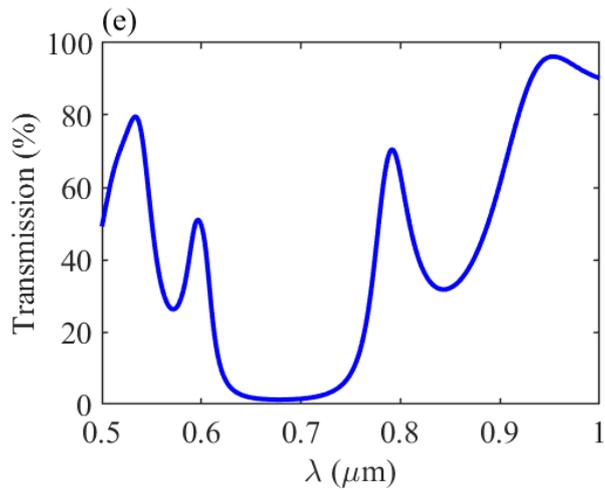
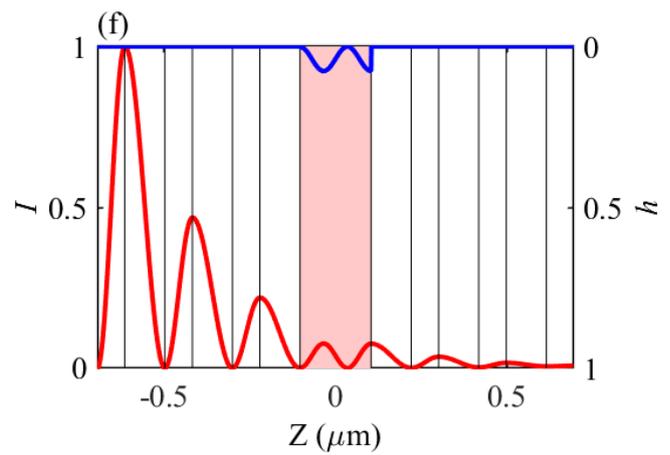

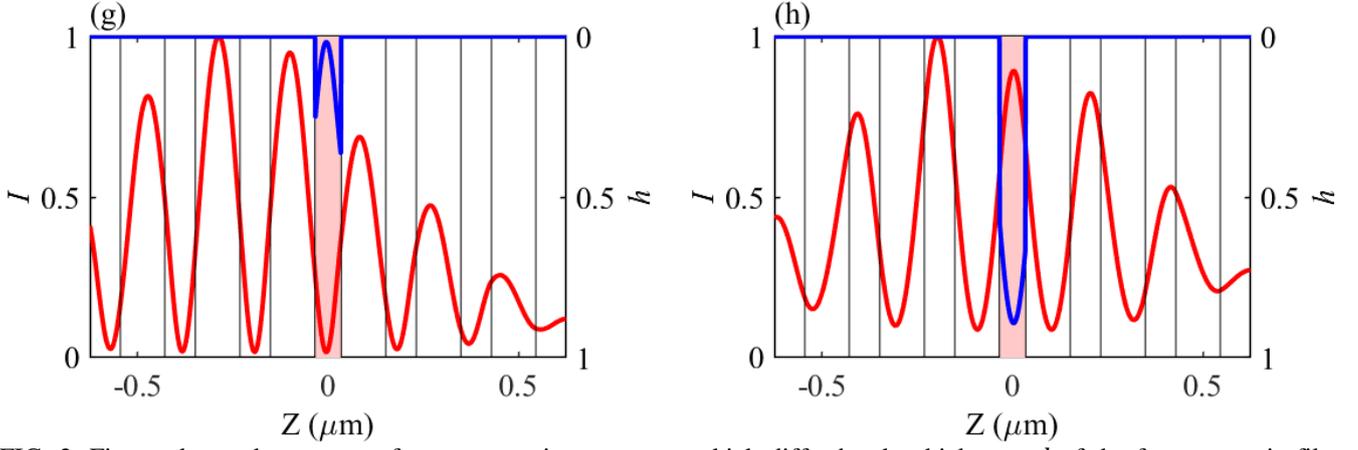

FIG. 2. Figure shows three cases of optomagnonic structures, which differ by the thickness $d$ of the ferromagnetic film located between the Bragg mirrors: $d = \frac{2\lambda_0}{4n} = 138 nm$ for (a) and (b); $d = \frac{\lambda_0}{4n} = 69 nm$ for (c), (d), (g), (h); $d = \frac{3\lambda_0}{4n} = 207 nm$ for (e), (f). Panels (a), (c), (e) show the transmission spectra of the optomagnonic structures. Panels (b), (d), (f), (g), (h) show $I = (E, E^*)$ distribution field (red curve) which indicates the light intensity inside the structure, and $h = Im\{[E \times E^*]_z\}$ -distribution (blue curve) which is responsible for the inverse Faraday effect inside the ferromagnetic film at wavelengths of 680 nm for (b), (d), (f), 580 nm for (g) and 840 nm for (h). In (d) the red arrow indicates the ordinate axis belonging to the left scale, the blue arrow - to the right scale. This also matches for (b), (f), (g), (h).

## V. SPECTRUM OF PSSW MODES EXCITED IN THE OPTOMAGNONIC CAVITY

In this section we will analyze the excitation of PSSW modes for different cases of the optomagnonic structures considered in Sec.IV. Variation of the magnetic layer thickness $d$ changes not only optical properties of the cavity but also its magnetic ones: the PSSW spectrum implicitly depends on the product of $\xi d$ through the wavevectors $k_n, \chi_{+,n}, \chi_n$ [see Eqs. (7), (8), Appendix (A7), (A8), (A14)–(A16)]. With the increase of $\xi d$ the modes change their form on the boundaries mostly due to the increase of the hyperbolic (surface) terms. Nevertheless, for $\xi d$, changing from 0.5 to 1.5, the impact of the surface terms is quite small, as it will be shown below.

Since in all considered cases $h(z)$ has a form of a constant with added harmonic function the spectrum of the excited PSSWs contains mostly zeroth and $s$-th harmonics (Figs. 3(b), 3(e), 3(h)). Here, the excitation amplitudes $A_n$ are calculated using Eqs. (10a)–(10b) assuming the following parameters of the magnetic film: $H = 1000$ Oe, $4\pi M = 1000$ G, $K_U = 10^3 \frac{erg}{cm^3}$, $A = 1.26 \cdot 10^{-10}$ cm$^2$, $K_s = 2.9 \cdot 10^{-2} \frac{erg}{cm^2}$, $\xi = \frac{2K_s}{AM^2} = 0.725 \cdot 10^5$ cm$^{-1}$, $\gamma = 1.76 \cdot 10^7 \frac{Hz}{Oe}$.

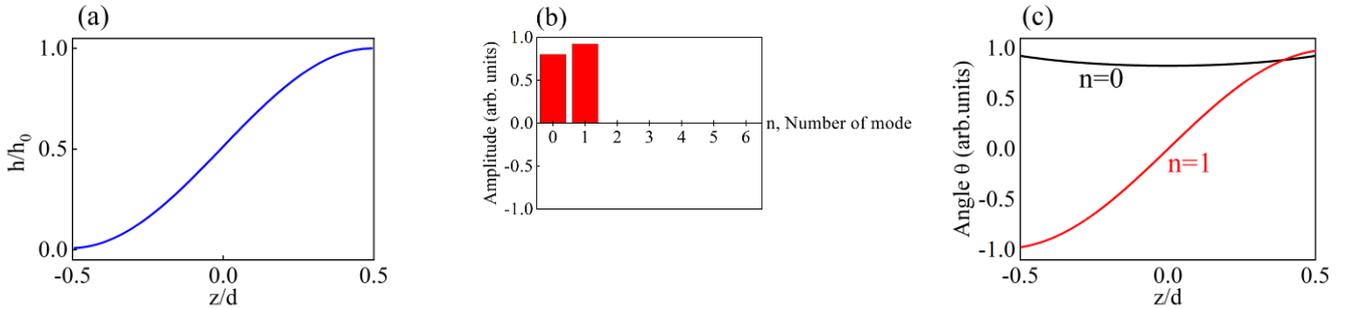

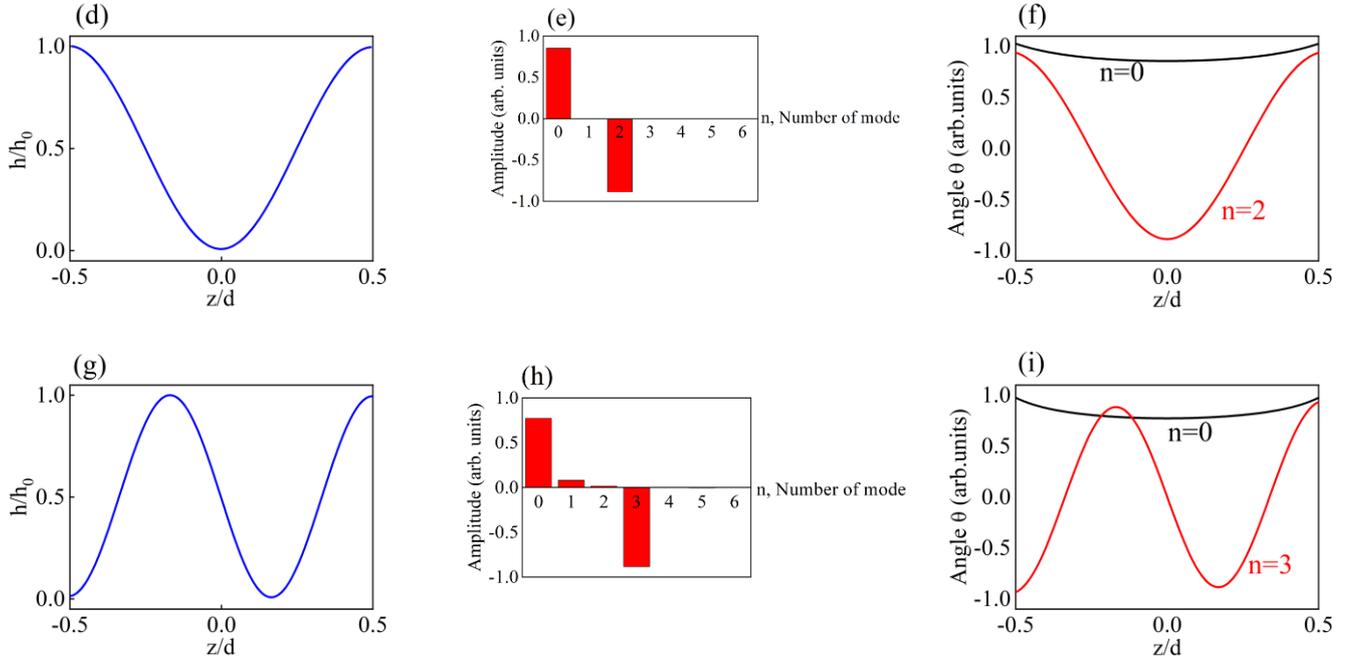

FIG. 3. Amplitudes $A_n$ of the PSSW modes (second column, (b), (e), (h)) for different IFE-field distributions (first column, (a), (d), (g)), corresponding to the different cases of optomagnonic structures described in FIG.2. IFE-field distributions may be described as follows: (a) $h(z) = \frac{h_0}{2}\left(1 + \sin\frac{\pi}{d}z\right)$; (d) $h(z) = \frac{h_0}{2}\left(1 - \cos\frac{2\pi}{d}z\right)$; (g) $h(z) = \frac{h_0}{2}\left(1 - \sin\frac{3\pi}{d}z\right)$. The third column (c), (f), (i) represents the form of two main eigenmodes $\theta_n(z)$, which are excited for the given configuration of the IFE-field.

Maximum amplitudes $A_n$ correspond to the modes $\theta_n(z)$, which have z-dependence close to $h(z)$ (compare $h(z)$ in Figs. 3(a), 3(d), 3(g) and corresponding $\theta_n(z)$ in Figs. 3(c), 3(f), 3(i) calculated through Eq. (7)). Thus for the structure with the quarter-wavelength magnetic layer the modes with $n=0$ and $n=1$ are mostly pronounced, while for the two times thicker layer $n=0$ and $n=2$ and for the magnetic layer of thickness $d = 3\lambda_0/4\sqrt{\varepsilon_m}$ the 0$^{th}$ and 3$^{rd}$ modes are more favorable. Though the fundamental mode with almost uniform spatial distribution ($n=0$) appears in all cases, the other modes with non-uniform distribution are excited for some particular thicknesses of the magnetic layer, which is quite important for their selective excitation.

We should also mention about the influence of $\xi d$. As $\xi d$ increases from 0.5 to 1.5 the boundary conditions (5) at the film surfaces, written in terms of the dimensionless coordinate $z' = z/d$, become far from the «free edges» ($\frac{\partial \theta}{\partial z'}|_{z'=\pm 1/2} \neq 0$; $\frac{\partial \theta}{\partial z'}|_{z'=\pm 1/2} = \pm \xi d \cdot \theta$) and therefore the behavior of the eigenmodes $\theta_n(z)$ deviates more from harmonic-like. It is mostly visible for the fundamental mode, whose distribution turns from almost constant (Fig. 3(c)) to the hyperbolic one (Fig. 3(i)), having its maximum amplitude at the edges. As a result, for the third case not only two basic modes ($n = 0,3$) are excited, but the other modes appear as well ($n = 1,2$), though their amplitudes remain relatively small. Consequently, one has to take into account that for the excitation of a single PSSW mode the product $\xi d$ should be not too large.

## VI. CONCLUSION

In this paper we suggested a one-dimensional optomagnonic structure enabling to effectively excite perpendicular standing spin waves by femtosecond laser pulses via the inverse Faraday effect. We found the exact form of PSSW modes [see Eq. (7)], which can be excited in a ferromagnetic film inside the optomagnonic microcavity: the «quasi-harmonic» and «hyperbolic» modes. The optomagnonic microcavity makes optical field distribution in the magnetic layer strongly non-uniform which provides excitation of different PSSW modes. Importantly, by a proper adjustment of the pulse wavelength one can excite either single PSSW mode or some superposition of the modes.

The other advantage of the suggested approach is the possibility to selectively excite PSSW modes of high orders, which is quite difficult to achieve using standard methods, as they do not provide strongly non-uniform distribution of microwave magnetic field in comparison with optomagnonic microcavities. Also, as the mechanism of excitation is the inverse Faraday effect, which is of non-thermal character, it won't lead to extra heat losses, which might be important for many applications. Moreover, the selectivity of PSSWs excitation looks promising for many novel applications in the field of quantum computing based on magnonic logic. The quantum technologies are developing very actively these days, and a precise optical control of spin dynamics is vital for implementing optical-to-microwave transducers, interfacing of the superconducting qubits, etc.

## ACKNOWLEDGEMENTS

This work was supported by the Russian Ministry of Education and Science, Megagrant project N 075-15-2019-1934.

## APPENDIX

### 1. PSSW modes

The dependence $\omega(k)$ is obtained in Eq. (6). Using it, we can find the inverse dependence $k(\omega)$ expressing $k$ as a root of quadratic equation:

$$k_\pm^2 = \frac{-[\widetilde{H} + H] \pm \sqrt{[\widetilde{H} - H]^2 + 4\frac{\omega^2}{\gamma^2}}}{2AM}. \quad (A1)$$

We see that two values of wavevector $k$ ($k_+$ and $k_-$) correspond to the given value of the frequency $\omega$. The first one ($k_+$) can be either real or imaginary, depending on $\omega$. If $\omega > \gamma\sqrt{\widetilde{H}H}$, then $k_+$ is real ($k_+^2 > 0$) and will be denoted simply as $k_+ = k$. In a case, when $\omega < \gamma\sqrt{\widetilde{H}H}$, $k_+$ becomes imaginary ($k_+^2 < 0$) and will be denoted as $k_+ = i\chi_+$. As for the second one ($k_-$), it is always imaginary ($k_-^2 < 0$) for any $\omega$ and, thus, can be represented as $k_- = i\chi$. Here $\chi_+, \chi$ are the real numbers ($\chi_+, \chi > 0$) As it will be shown later, modes with imaginary $k_\pm$ (represented by $\chi_+, \chi$) are responsible for hyperbolic (surface) terms of modes, and can't be neglected. To sum it up, two different cases are possible:

$$k_+ = k \, ; \, k_- = i\chi \quad (k_+^2 > 0 \, ; \, k_-^2 < 0), \quad (A2.a)$$

$$k_+ = i\chi_+ \, ; \, k_- = i\chi \quad (k_+^2 < 0 \, ; \, k_-^2 < 0). \quad (A2.b)$$

From Eq. (A1) it can be shown that different wavevectors ($k_+$ and $k_-$) are linked together by the ratio:

$$\chi^2 = k^2 + \frac{\widetilde{H} + H}{AM}, \quad k_+^2 > 0, \quad (A3.a)$$

$$\chi^2 = -\chi_+^2 + \frac{\widetilde{H} + H}{AM}, \quad k_+^2 < 0. \quad (A3.b)$$

Since the boundary conditions (5) are symmetric (in the sense that $\xi$ is the same on the boundaries), we will be searching for solutions, symmetric or antisymmetric with respect to z-coordinate.

$$\varphi(z), \theta(z) \sim \begin{cases} \cos k_+ z + B \cosh \chi z \\ \sin k_+ z + B \sinh \chi z \end{cases}, \quad (A4)$$

where $B$ coefficients will be determined later.

We should add the hyperbolic terms in Eq. (A4), as the boundary conditions (5) for $\theta$ and $\varphi$ are different and using only harmonic functions will not be sufficient for the satisfaction of all boundary conditions.

a) **Case of «quasi-harmonic» modes ($k_+^2 > 0$):**

From Eqs. (3), (4) we see that $\varphi(t) \sim e^{-\lambda t} \cos \omega t$, $\theta(t) \sim e^{-\lambda t} \sin \omega t$. As for the z-dependence $\varphi(z), \theta(z)$, it is described by Eq. (A4). Then, in the case, when $k_+^2 > 0$ ($k_+ = k$), the solution for symm./antisym. modes finally takes the form:

$$\varphi(z,t) \sim \begin{cases} \cos kz + B_\chi \cosh \chi z \\ \sin kz + B_\chi \sinh \chi z \end{cases} \cdot e^{-\lambda t} \cos \omega t, \qquad (A5)$$

$$\theta(z,t) \sim \begin{cases} \cos kz + b_\theta B_\chi \cosh \chi z \\ \sin kz + b_\theta B_\chi \sinh \chi z \end{cases} \cdot e^{-\lambda t} \sin \omega t. \qquad (A6)$$

From the second expression in Eq. (3), neglecting $\alpha$, we can find $b_\theta$:

$$b_\theta = -\frac{\widetilde{H} + AMk^2}{H + AMk^2}. \qquad (A7)$$

From the boundary condition on $\varphi$ [Eq. (5)], we can find $B_\chi$:

$$B_\chi = \begin{cases} \dfrac{k}{\chi} \dfrac{\sin\frac{kd}{2}}{\sinh\frac{\chi d}{2}}, & \text{for symm. modes} \\[2mm] -\dfrac{k}{\chi} \dfrac{\cos\frac{kd}{2}}{\cosh\frac{\chi d}{2}}, & \text{for antisymm. modes} \end{cases} \qquad (A8)$$

We should mention, that for small values of $\xi d$ (which is our case, $\xi d = 0.5 - 1.5$) hyperbolic terms in Eqs. (A5), (A6) are also small, hence, the modes described by Eqs. (A5), (A6) are almost harmonic. We will call them «quasi-harmonic» modes.

Now let us find the dependence of wavevector $k_+$ on the number of mode $n$ (in a case $k_+^2 > 0$). The modes will be numbered starting from zero mode ($n = 0,1,2 ...$). Using the boundary condition on $\theta$ [Eq. (5)] and substituting symmetric and antisymmetric solutions [Eq. (A6)] in it, we will have the expressions, which implicitly determine $k(n)$:

$$\begin{cases} \left[1 + \dfrac{\xi}{k}\cot\dfrac{kd}{2}\right] = b_\theta \left[1 - \dfrac{\xi}{\chi}\coth\dfrac{\chi d}{2}\right], & \text{for symm. modes} \\[2mm] \left[1 - \dfrac{\xi}{k}\tan\dfrac{kd}{2}\right] = b_\theta \left[1 - \dfrac{\xi}{\chi}\tanh\dfrac{\chi d}{2}\right], & \text{for antisymm. modes} \end{cases} \qquad (A9)$$

If $\frac{\xi}{\chi} \ll 1$; $\chi d > 4$ (which is valid for the given parameters), then, we can simplify the Eqs. (A9):

$$\begin{cases} -\cot\dfrac{kd}{2} = \dfrac{k}{\xi^*}, & \text{for symm. modes} \\[2mm] \tan\dfrac{kd}{2} = \dfrac{k}{\xi^*}, & \text{for antisymm. modes} \end{cases} \qquad (A10)$$

Here $\xi^* = \xi \cdot \frac{1}{1-b_\theta} = \xi \cdot \frac{H+AMk^2}{\widetilde{H}+H+2AMk^2}$. For large values of $k$ we have: $\xi^* \to \frac{\xi}{2}$.

Let us find an approximate analytical solution of Eqs. (A10), which is valid with high accuracy, if $\frac{\xi^* d}{\pi n} < 1$. Then:

$$k_n = \frac{\pi n}{d} - \Delta k = \frac{\pi n}{d} - \frac{\pi n - \sqrt{\pi^2 n^2 - 8\xi^* d \left(1 - \frac{\xi^* d}{6}\right)}}{2d\left(1 - \frac{\xi^* d}{6}\right)}. \qquad (A11)$$

If $\frac{8\xi^* d \cdot \left(1-\frac{\xi^* d}{6}\right)}{\pi^2 n^2} \ll 1$, then we can simplify Eq. (A11) and, thus, obtain $k(n)$ dependence described by Eq. (8). Note, that the zero mode with $n = 0$ is not mentioned in Eq. (8), as it does not satisfy the condition $k_+^2 > 0$ and will be described in the next paragraph.

b) **Case of «hyperbolic» modes ($k_+^2 < 0$):**

We should also analyze the case, when $k_+^2 < 0$ ($k_+ = i\chi_+$). This is possible for the first two modes ($n = 0,1$). Then, the expressions for these modes are the following:

$$\varphi(z,t) \sim \begin{cases} \cosh \chi_+ z + \widetilde{B}_\chi \cosh \chi z \\ \sinh \chi_+ z + \widetilde{B}_\chi \sinh \chi z \end{cases} \cdot e^{-\lambda t} \cos \omega t, \qquad (A12)$$

$$\theta(z,t) \sim \begin{cases} \cosh \chi_+ z + \tilde{b}_\theta \tilde{B}_\chi \cosh \chi z \\ \sinh \chi_+ z + \tilde{b}_\theta \tilde{B}_\chi \sinh \chi z \end{cases} \cdot e^{-\lambda t} \sin \omega t, \quad (A13)$$

$$\tilde{b}_\theta = -\frac{\tilde{H} - AM\chi_+^2}{H - AM\chi_+^2}, \quad (A14)$$

$$\tilde{B}_\chi = \begin{cases} -\dfrac{\chi_+}{\chi} \dfrac{\sinh \frac{\chi_+ d}{2}}{\sinh \frac{\chi d}{2}}, & \text{for symm. mode} \\ \\ -\dfrac{\chi_+}{\chi} \dfrac{\cosh \frac{\chi_+ d}{2}}{\cosh \frac{\chi d}{2}}, & \text{for antisymm. mode} \end{cases} \quad (A15)$$

The modes described by Eqs. (A12), (A13) represent a sum of two hyperbolic functions. We will call them «hyperbolic» modes. Here, a symmetric mode corresponds to $n = 0$, an antisymmetric – to $n = 1$.

For «hyperbolic» modes the dispersion relations (A9) are also changing:

$$\begin{cases} \left[1 - \dfrac{\xi}{\chi_+} \coth \dfrac{\chi_+ d}{2}\right] = \tilde{b}_\theta \left[1 - \dfrac{\xi}{\chi} \coth \dfrac{\chi d}{2}\right], & n = 0, \\ \left[1 - \dfrac{\xi}{\chi_+} \tanh \dfrac{\chi_+ d}{2}\right] = \tilde{b}_\theta \left[1 - \dfrac{\xi}{\chi} \tanh \dfrac{\chi d}{2}\right], & n = 1. \end{cases} \quad (A16)$$

The wavevectors $\chi_+$, $\chi$, which are linked together by the ratio (A3.b), can be found from Eqs. (A16) by solving them numerically. However, for the small values of product $\xi d$ ($\xi d < 0.5 - 1.0$) the first expression in Eqs. (A16) (responsible for $n = 0$ mode) may be simplified and we can deduce an approximate formula for $\chi_{+,0}$ depending on $\xi d$: $\chi_{+,0} d = \sqrt{\frac{2}{3} \xi d}$.

For some values of product $\xi d$ «hyperbolic» modes $n = 0,1$ may not exist. The exact conditions for the existence of «hyperbolic» modes can be found from the following considerations. Firstly, the wavevector $k_+$ should be imaginary ($k_+^2 < 0$), otherwise, «hyperbolic» modes will turn to «quasi-harmonic» modes. Secondly, the oscillation frequency should be real ($\omega^2 > 0$), otherwise, modes cannot exist. All this leads to the following condition on $\chi_+$:

$$0 < \chi_+ < \sqrt{\frac{H}{AM}}. \quad (A17)$$

The wavevector $\chi_{+,n}$ depends on the number of mode ($n = 0,1$) and the value of dimensionless product $\xi d$. Thus, using dispersion relations (A16), the condition (A17) may be reformulated for the first two modes in terms of $\xi d$:

$$0 < \xi d < \frac{\sqrt{\frac{\tilde{H}}{AM}} d}{\coth \left[\sqrt{\frac{\tilde{H}}{AM}} \frac{d}{2}\right]}, \quad n = 0, \quad (A18)$$

$$6 \cdot \left(1 + \frac{4 \tanh \left[\sqrt{\frac{\tilde{H} + H}{AM}} \frac{d}{2}\right]}{\sqrt{\frac{\tilde{H} + H}{AM}} d}\right)^{-1} < \xi d < \frac{\sqrt{\frac{\tilde{H}}{AM}} d}{\tanh \left[\sqrt{\frac{\tilde{H}}{AM}} \frac{d}{2}\right]}, \quad n = 1. \quad (A19)$$

We see that the zero mode $n = 0$ is always «hyperbolic» [Eq. (A18)] or does not exist for large values of $\xi d$: $\xi d > \frac{\sqrt{\frac{\tilde{H}}{AM}} d}{\coth \left[\sqrt{\frac{\tilde{H}}{AM}} \frac{d}{2}\right]}$. As for the first mode $n = 1$, it is «quasi-harmonic» ($k_+^2 > 0$) for small values of $\xi d$: $0 < \xi d < 6 \cdot \left(1 + \frac{4 \tanh \left[\sqrt{\frac{\tilde{H}+H}{AM}} \frac{d}{2}\right]}{\sqrt{\frac{\tilde{H}+H}{AM}} d}\right)^{-1}$. For larger $\xi d$ the mode becomes «hyperbolic» [Eq. (A19)]. And finally, with the further increase of $\xi d$ the mode disappears.

Let us now make some estimations. For $d = 70$ nm the conditions (A18), (A19) on $\xi d$ are the following: $0 < \xi d < 2.86$ for $n = 0$ and $3.00 < \xi d < 3.41$ for $n = 1$. To compare with, if we take $d = 140$ nm, we will have: $0 < \xi d < 6.46$ for $n = 0$ and $3.99 < \xi d < 6.50$ for $n = 1$. Thus, we see that in a case, when $\xi d = 0.5 - 1.5$, both «hyperbolic» modes exist and the first mode $n = 1$ is a «quasi-harmonic» mode.

## 2. Excitation amplitudes for different PSSW modes

The general solution $\theta(z,t)$ will be the sum of different PSSW modes $\theta_n(z,t)$ [Eq. (7)] with coefficients $A_n$ [Eq. (9)]. To find coefficients $A_n$, which are the excitation amplitudes for different eigenmodes, we will use the initial condition on time-derivatives $(\dot\theta, \dot\varphi)$, which can be obtained by substituting Eq. (4) in Eq. (3):

$$\begin{cases} \dot\varphi(z, t = \Delta t) = -\alpha\gamma^2 \Delta t\left[(H + \widetilde{H}) \cdot h(z) - 2AM \cdot h''(z)\right] \cong 0, \\ \dot\theta(z, t = \Delta t) = \gamma^2 \Delta t[H \cdot h(z) - AM \cdot h''(z)]. \end{cases} \quad (A20)$$

Substituting Eq. (9) in a condition on $\dot\theta$ [Eq. (A20)], we will have:

$$\dot\theta(z, t = \Delta t) = \sum_{n=0}^{\infty} A_n \omega_n \theta_n(z) = \gamma^2 \Delta t[H \cdot h(z) - AM \cdot h''(z)]. \quad (A21)$$

In a real experiment the IFE-field may be proportional to the superposition of the harmonic functions $h_s(z) \sim \sin k_s z, \cos k_s z$ (including $k_s = 0$): $h(z) = \sum_s h_s(z)$. In this case, the Eq. (A21) may be simplified using $h_s''(z) = -k_s^2 \cdot h_s(z)$ :

$$\sum_{n=0}^{\infty} A_n \omega_n \theta_n(z) = \gamma^2 \Delta t \sum_s [H + AMk_s^2] \cdot h_s(z). \quad (A22)$$

For the small values of $\xi d$ ($\xi d < 1.5 - 2.0$) the eigenmodes are almost orthogonal: $\int_{-\frac{d}{2}}^{\frac{d}{2}} \theta_m(z)\theta_n(z)\, dz \approx \frac{d}{\beta_n}\delta_{mn}$ ($\beta_{n=0} = 1, \beta_{n \neq 0} = 2$). Then, multiplying Eq. (A22) by $\theta_m(z)$ and integrating it by $z$ with the further change of index $m$ by $n$, we will obtain the expression for the amplitudes of modes $A_n$:

$$A_n = \frac{\beta_n \gamma^2 \Delta t}{\omega_n d} \sum_s (H + AMk_s^2) \int_{-d/2}^{d/2} h_s(z) \cdot \theta_n(z)\, dz. \quad (A23)$$

In Eq. (A23) the frequency $\omega_n$ may be expressed through the wavevector $k_n$ or wavevector $\chi_{+,n}$ [see Eqs. (6), (A2.a), (A2.b)], depending on the character of mode $\theta_n(z)$ («quasi-harmonic» or «hyperbolic»). Then, we will finally have Eqs. (10.a)–(10.b).

Thus, by creating different IFE-field distributions $h(z)$, we can excite different spectra of modes. Let us consider several examples of harmonic IFE-field distribution $h_s(z)$ with $k_s = s\pi/d$ ($s$ is odd or even):

a) **IFE-fields with «free edges»** ($h_s'(z) = 0$ on the edges): $h_s(z) = \cos\left(\frac{s_{even}\pi z}{d}\right), \sin\left(\frac{s_{odd}\pi z}{d}\right)$.

These IFE-fields mostly excite the single mode with the number $n = s$.

b) **IFE-fields with «fixed edges»** ($h_s(z) = 0$ on the edges): $h_s(z) = \cos\left(\frac{s_{odd}\pi z}{d}\right), \sin\left(\frac{s_{even}\pi z}{d}\right)$.

These IFE-fields excite the spectrum of modes, while the maximum value of amplitude $A_n$ corresponds to the mode with the number $n = s - 1$.

///references///